\documentstyle[12pt]{article}
\date{}
\newcommand{\bi}{\bibitem}
\newcommand{\be}{\begin{equation}}
\newcommand{\ee}{\end{equation}}
\newcommand{\ba}{\begin{eqnarray}}
\newcommand{\ea}{\end{eqnarray}}

\begin{document}
\noindent
{\Large\bf Fine tuning in the\
Standard Model and Beyond\footnote{Talk given at the X Workshop on
Quantum Field Theory and High Energy Physics, September 1995, Zvenigorod,
Russia}}\\
\vspace{5mm}

A. A. Andrianov$^{1}$, \quad N. V. Romanenko$^2$\\

\medskip

\noindent
{\small$^1$
 Institute of Physics,
University of Sankt-Petersburg,
198904 Sankt-Petersburg, Russia\\
$^2$ Department of  Neutron Research,
Petersburg Nuclear  Physics Institute,
 188350, Gatchina, Russia}

\vspace{1cm}

\noindent
{\bf Abstract.} The fine-tuning principles are examined to
predict the top-quark and Higgs-boson masses. The modification of the
Veltman condition based on the
compensation of vacuum energies
 is developed. It is implemented in the Standard Model and in its minimal
 extension with two Higgs doublets and Left-Right symmetric Model.
The top-quark and Higgs-boson couplings are fitted in the SM for
the lowest ultraviolet scale where the fine-tuning can be stable under
rescaling. It yields
the low-energy values $m_t \simeq 175 GeV;\quad m_H \simeq 210 GeV$.
For the Two-Higgs and Left-Right Symmetric Models the fine-tuning
principles  yield the interval for top-quark mass, compatible
with the modern experimental data.
For the Left-Right Model the FT principles  demand the existence
of the right-handed Majorana neutrinos with masses of order of
right-handed gauge bosons.

\vspace{5mm}

\noindent
{\bf 1. \quad Introduction}\\

\medskip

The Standard Model (SM) describes
the strong and electroweak
particle interactions for a whole range of
energies which have been
available in experiments \cite{1}. Still, there is a number of
well-known problems that are
to be
resolved in order to justify all the principles which the Standard
Model is based on. In particular, the precise top quark mass
measurement is wanted  \cite{1,2}
the discovery of scalar Higgs particle is desired
\cite{1,3}.

There exist a few phenomenological principles
within the SM and its minimal extensions which make it possible
to determine relations between top-quark and
Higgs-boson masses. These principles are based on the assumption
that the SM is actually an effective theory applicable for
low energies. Let us formulate them as follows:
\begin{itemize}
\item The strong fine tuning for the Higgs field parameters
(v.e.v and its mass) that consists in the
cancellation of large radiative contributions quadratic in ultraviolet scales
which bound the particle spectra in the effective theory (Veltman condition
\cite{8}-\cite{12}).
\item The strong fine tuning for vacuum energies \cite{13}
that envisages the cancellation of
large divergencies quartic in ultraviolet scales which might effect
drastically the formation of the cosmological constant.
\item The RG stability of the cancellation mechanism under
change of ultraviolet scale of effective theory \cite{10,13}.
\end{itemize}

In our paper we examine the compatibility of these principles
for the SM,  for the two-Higgs model and the Left-Right
Symmetric model.

\bigskip

{\bf 2.\quad Vacuum fine-tuning in the Standard Model}\\

\medskip

Let us consider the SM as a low-energy limit of a more
fundamental theory and suppose that only one heavy fermion, t-quark
takes part in its dynamics within the selected energy range. Respectively,
 we
neglect the masses of all lighter fermions. We apply different
scales for the design of SM-effective action for bosons $\Lambda_{B}<
\Lambda_{new}$ and for fermions $\Lambda_{F}< \Lambda_{new}$.

Strong fine-tuning for vacuum energies reads at one loop level:
\be
\alpha^2 = \frac{\Lambda_B^4}{\Lambda_F^4}=
 \frac{45}{14} \quad(\mbox{or}\quad \frac{24}{7});
\ee
The case in brackets intends that the neutrinos are
 massive Dirac particles.

Respectively the strong fine-tuning condition
for the Higgs parameters (at the one-loop
level) reads:
\be
f \equiv 4m^2_t-\alpha(2M^2_W+M^2_Z+m^2_H)=0;\quad \alpha \simeq 1.793
\quad(\mbox{or}\quad 1.852). \label{eq:xx}
\ee

 Let us prolongate the validity of the fine tuning to the entire
 energy range below i.e. demand the RG stability
 (or weak dependence of scale) for the abovementioned conditions:
 \be
Df \equiv 16 \pi^2 \frac{\partial f}{\partial \tau} =0;\quad
\tau = \ln\frac{\Lambda}{v_0}. \label{MV1}
\ee

It can be found  that the solution  for $g_t^2$
exists only for the effective scale $\Lambda \sim 10^{15} GeV$
when EW coupling
constants approach to their GUT values $g^2_3 \sim g^2 \sim (5/3)g'^2$.
The low-energy value for the
Yukawa coupling constant $g_t$ is delivered by the renormalization-group
flow.

The low energy values of $m_H$ are obtained with the help of the IR
quasi-fixed
point in the RG-equation for the Higgs self-coupling \cite{16}

The stability condition ensures the strong f\/ine-tuning both
to two-loop level and numerically and leads at the EW scale to the
predictions,
\ba
\left\{
\begin{array}{l}
m_t (v) = 175 \pm 5\, GeV,\\
m_H (v) = 210 \pm 10\, GeV.
\end{array} \right.
\ea
The dependence of the neutrino degrees of freedom is rather weak
and is included into the error bar.

One can check up  that the modified Veltman equation does not
depend on the rescaling for the wide range of energies.

For more detailed discussion see \cite{13}.

\bigskip

\noindent
{\bf  3.\quad Vacuum fine-tuning in the Two-Higgs Model}\\

\bigskip

We consider the Two-Higgs model with the
most general potential
which possesses the discrete symmetry,
 $H_1 \rightarrow H_1$; $H_2 \rightarrow -H_2$ \cite{4}.

 The constants in the Higgs potential should make it bounded from below and
 should
 guarantee that the vacuum
 configuration conserve electric charge
 ($\lambda_4<0$). The choice of the $H_2$ phase may be always done so that
 $\lambda_5<0$.

The realistic ratio $m_b/m_t << 1$ can be produced in several ways.\\
1) The mass hierarchy can be treated as the consequence
of a hierarchy of coupling constants,
$ g_b <<  g_t,$ while both
v.e.v.'s are comparable in their magnitudes.
One can check up that in this case the cancellation of the
quadratic divergences cannot be supplied with the one-loop RG invariance
at every scale \cite{13}.\\
2)The mass difference may be caused by the hierarchy of v.e.v's:
$m_b/m_t= v_2/ v_1 << 1$,
 $g_b= g_t \equiv F;$
For such a choice the global
right symmetry arises in the limit $g' \rightarrow 0$.
Then it happens to be possible  to
cancel both vacuum energies  and quadratic divergencies and
furthermore to implement the one-loop RG invariance of these conditions.\\
3) More general model seems to be less natural since
it is difficult to avoid non-conserving strangeness
neutral currents when the Cabibbo mixing is taken into account.
By this reason
we restrict ourselves with the analysis of second case.\\

Let us derive the vacuum fine-tuning conditions for the Two-Higgs model.
The vacuum-energy cancellation  reads:
$$
\alpha \equiv \frac{\Lambda_B^2}{\Lambda^2_F}
\approx 1.677 \quad(\mbox{or}\,\,\, 1.732).$$

In this case the fine-tuning conditions for Higgs parameters
and its RG invariance:
  \be \left\{ \begin{array}{l}
  f\equiv 4F^2- \alpha[\frac{3}{2}g^2+\frac{1}{2}g'^2+
  \frac{10}{3}\bar\lambda+\frac{2}{3}\lambda_4]=0; \nonumber \\
  Df=0;
  \end{array} \right. \ee

can be reduced to the equation for the Yukawa coupling constant
when excluding $\bar\lambda$ from the first equation.
It can be seen that this system has one
positive solution for any constants $g_3, g, g', \lambda_{4,5}$.
Below on the numerical estimations
 of minimal values for $m_t$ (this corresponds to
 $\lambda_{4,5} \simeq0$ )
 are presented for different energies.

In order to predict the real $m_t$ we use the RG flow
Predictions for the Higgs spectra can be found with help of
the quasi IR fixed points [17].
For the chosen scheme of couplings:
$m_+ \approx 200-205 \;GeV $; $m_1 \approx 225-230 \; GeV$;
$m_2 \approx 6-6.5 \;GeV  $; $m_p \approx 0\; GeV$ (PQ-symmetry).

It is surprising that the estimations for
the $t$-quark mass are close to the fine-tuning predictions
of the one-Higgs Standard Model and to the recent experimental
data \cite{2}.\\

\begin{center}

{\bf Table 1.}\quad Masses of the $t$-quark for $\lambda_{4,5} \simeq0$.

\begin{tabular}{||l|c|c|c|c||} \hline \hline
$\Lambda \, GeV$
& $``m_t(\Lambda,\nu_{Weyl})"$ &$m_t(100 \, GeV, \nu_{Weyl})$&
$ ``m_t(\Lambda,\nu_{Dir})"$
&$m_t(100 \, GeV, \nu_{Dir})$ \\ \hline
$10^{15}$ &  101& 162     &  99 & 160  \\ \hline
$10^{14}$ &  104& 163     &   101 & 161 \\ \hline
$10^{13}$ &  108& 165     & 104   & 162 \\ \hline
$10^{12}$ &  112& 167     & 108   & 164 \\ \hline
$10^{11}$ &  119& 170      & 113   & 166 \\ \hline
$10^{10}$ &  127& 174      & 120   & 170 \\ \hline
$10^{9}$  &  140& 181      & 129   & 175 \\ \hline
$10^{8}$  &  163& 194      & 143   & 182 \\ \hline
$10^{7}$  &  212& 216      & 165   & 195 \\ \hline\hline
\end{tabular}

\end{center}

\bigskip

\noindent
{\bf  4.\quad Vacuum fine-tuning in the Left-Right Symmetric Model}\\

\medskip

  The model with Left-Right gauge symmetry
  ( $SU(3)_c*SU(2)_L*SU(2)_R*U(1_Y)$)
   is widely discussed
as a possible candidate for   generalization of the Standard Model.
It is rather attractive both for theoretical reasons
and for experimental ones as well ---connected with some discrepancies
in the precise experimental values within the framework of the
Standar Model \cite{lr}.

Theory contains three usual generations of Standard Model fermions
with the obligatory addition of right-handed neutrinos
(detailed discussion of their quantum numbers see in \cite{lr}).
The gauge sector differs from the SM by the obvious addition of the
right-handed gauge bosons. Besides that the Higgs sector of the model
contains much more particles than in the SM \cite{3}.

For the generation of the fermion masses one with necessity needs
the Higgs bidoublet with the following quantum numbers
$(T_L,\;T_R,\;Y)$:
$$\Phi=\left(
\begin{array}{ll}
\phi_1^0 & \phi_1^+ \nonumber \\
\phi_2^- & \phi_2^0
\end{array}
\right)=(\frac{1}{2}, \frac{1}{2}^*,0).
$$
This field  has to acquire nonzero vacuum expectation,
saving however the electromagnetic invariance of vacuum
(i.e only for neutral components).
The existence of the abovementioned bidoublet is not enough to yield
the spontaneous symmetry breaking of the $SU(2)_L*SU(2)_R$ gauge group.
 Additional Higgs choice is alternative:

 a) Higgs doublets:
 \be
 \delta_L=\left[
 \begin{array}{l}
 \delta_L^+  \\
 \delta_L^0
 \end{array}
 \right]=(\frac{1}{2},0 ,1); \;
 \delta_R=\left[
 \begin{array}{l}
 \delta_R^+  \\
 \delta_R^0
 \end{array}
 \right]=(0, \frac{1}{2},1)
 \ee

They can generate heavy right-handed gauge bosons, but cannot
interact with fermions.

b) Higgs triplets:
\be
\vec{\Delta}_L= \left[
\begin{array}{l}
\Delta_L^{++} \\
\Delta_L^+ \\
\Delta_L^0
\end{array}
\right]=(1,0,2); \;\;\;
\vec{\Delta}_R= \left[
\begin{array}{l}
\Delta_R^{++} \\
\Delta_R^+ \\
\Delta_R^0
\end{array}
\right]=(0,1,2)
\ee

They can produce large $M_{W_R}$ as well as Majorana masses for
neutrino.

These facts lead to an obvious observation: since the
  cancellation of all  quadratic divergencies
in the scalar sector occurs due to the
compensation between the bosonic and fermionic loops
 the case a) should be abolished,
because no fermions with usual quantum numbers can
have Yukawa couplings to scalar doublets.
In the case b) with the triplet Majorana-Higgs  representation
  Majorana masses for right-handed neutrinos should be
  generated,
having the same order of magnitude as right-handed
gauge bosons. Then one can expect the cancellation of the quadratic
divergences in the scalar sector. The presence of the $\Delta_L$
fields in general case is not necessary. But if these fields exist
-- for example in the case of the manifest LR symmetry,
the vacuum expectation  of the left-handed Higgs-Majoron
should be  extremely small because of the upper bound
$\sim 1 eV$ on the left-handed Majorana neutrino masses.

Thus the FT in the LR Model leads to the unambiguous determination
of the symmetry breaking sector of the theory.

The condition of vacuum energy cancellation reads in this theory
as follows:

$$\alpha^2= \frac{4N_F}{2N_B+N_S}=
\left[
\begin{array}{l}
96/50 \;\; if\; \Delta_L  \; exist\\
96/44 \;\; without \; \Delta_L
\end{array}
\right]
$$

In order to obtain fine-tuning for the Higgs parameters
one needs the concrete form of the Higgs potential and
the Yukawa couplings.
For the same reasons as in the Two-Higgs Model we shall
 assume that the top-bottom mass difference is
 produced by the hierarchy of v.e.v's ($F\equiv g_t=g_b$).

   The general form of Higgs potential of such model
  contains  15 (!) self-couplings  \cite{lr}.

  We shall consider the
  situation, when   interactions between
  fields $\Phi$ and $\Delta_{L,R}$ are suppressed
  set to zero.
  Then the simplest
  renorm-invariant potential for the  $\Phi$  fields contains
  4                  self-couplings:
  $$V_{\Phi}\sim l_1 Tr^2(\Phi \Phi^+)+
  l_2[Tr^2(\tilde{\Phi} \Phi^+)+Tr^2(\tilde{\Phi^+}\Phi)]+
  l_3Tr(\tilde{\Phi}\Phi^+)Tr(\tilde{\Phi^+}\Phi)+$$
  $$+a_1 \left[ Tr(\Phi \Phi^+) \right]
   \cdot \left[
   Tr(\Delta_L \Delta_L^+)
    + Tr(\Delta_R \Delta_R^+) \right] $$

 The modified Veltman equations for  vacuum
  expectations of $\Phi$   read
  for the case without $\Delta_L$ (with $\Delta_L$):
  \be
    f_{\Phi}\equiv 4F^2-\alpha[\frac{3}{2}g_L^2+\frac{3}{2}g_R^2
  +\frac{20}{3}l_1 +\frac{8}{3}l_3+ 2a_1\; (+4a_1)]=0; \;
(  Here \;a_1 \approx 0)
      \ee

Adding to this the weak fine-tuning condition $Df_{\Phi}=0$
one can obtain estimations for the top-quark mass.

\vbox{
\begin{center}
{\bf Table 2.  Masses of the $t$-quark in the LR Model.}

{}~~~~~~~~Case  without  $\Delta_L$. ~~~~~~~~~~Case with $\Delta_L$.

\begin{tabular}{||l|c|c|c|c||} \hline \hline
$\Lambda \, GeV$ &  $``m_t(\Lambda)"$ &$m_t(100 \, GeV)$ &  $``m_t(\Lambda)"$
&$m_t(100 \, GeV)$ \\ \hline
$10^{15}$        &   106--287 &165--202  &  108--226 & 166--197           \\
\hline
$10^{14}$        &   109--287 &  166--205    &   111--225 & 167--199
\\ \hline
$10^{13}$        &  112--287   & 167--208   & 114--224   & 169--201          \\
\hline
$10^{12}$        &  115--286   & 169--211  & 118--222   & 170--204          \\
\hline
$10^{11}$        &  119--284   & 170--214  & 123--219   & 173--206          \\
\hline
$10^{10}$        &  124--280   & 172--218  & 130--214   & 176--208          \\
\hline
$10^{9}$         &  131--274   & 176--222  & 140--205   & 181--208          \\
\hline
$10^{8}$         &  139--265   & 180--226  & 159--185   & 192--209          \\
\hline
$10^{7}$         &  153--248   & 188--227  & ---   & ---            \\ \hline
$10^{6}$         &   ---       & ---       &  ---       & ---
\\ \hline
\end{tabular}
\end{center} }

The denotation $``m_t(\Lambda)"$  means $g_t(\lambda) \cdot 175$ GeV.
It can be easily checked up that the fine-tuning equations
 can have positive solutions
for only such values of gauge couplings which they
have at energies much more than $100 \, GeV$.
Let us assume that at the scale $\Lambda$ the left-handed
and the right-handed couplings are equal: $g_L=g_R$
(in the absence of $\Delta_L$ fields they have different RG-flows).
Then the solutions of the above equations result in rather
narrow range of possible values for $m_t$ for different
values of $\Lambda$. The   extreme values (the largest
and the smallest) of $m_t$ correspond to the choice
$l_3=l_2=0$, while the nonzero values
of these self-couplings push the two possible
values of the top mass inside the
interval  presented in the following tables.
One can see, that the above equations contain the restrictions
on the maximal possible values of the $l_3$ and $l_2$.
For the gauge couplings the experimental input was taken
from \cite{15}.
To predict the real $m_t(100\, GeV)$ we use RG flow.
The obtained  restrictions seem to be in a good agreement
with the modern experimental results of the top-quark
 search.

Now let us analyze the strong fine-tuning condition for
the Higgs-Majoron fields $\Delta_R$. It reads:
\be
2h_M^2= 3\alpha[2(2g^2+g'^2)+32\rho_1+16\rho_2]
\ee

Here $h_M$ is the Majorana-Yukawa coupling of the type:
 \be L_{MYu}\sim
 -\frac{h_M}{2}
 \overline{\omega}(i\tau_2 \Delta_R \frac{(1-\gamma_5)}{2}
 -\Delta_R^+i\tau_2\frac{(1+\gamma_5)}{2})\omega;
  \omega \equiv \psi_R +C\overline{\psi}_R^T; \ee
$$\Delta_R= \Delta_R^{++} \tau^+
+ \Delta_R^+ \tau^3 / \sqrt{2} + \Delta_R^0 \tau^-
$$

The $\rho_{1,2}$    are the constants
of the Higgs-Majoron potential (taking into account $a_1\approx 0$):
$$V_{\Delta} \sim
\rho_1 tr^2(\Delta_R \Delta_R^+)
+\rho_2 tr(\Delta_R)^2tr(\Delta_R^+)^2 $$

 From positive-definiteness of the Higgs-Majoron potential one can
get that $\rho_1>0$, $\rho_1+\rho_2>0$. Then from the modified
Veltman condition comes:
\be
2h_M^2 \ge 6\alpha(2g^2+g'^2)
\ee
 Taking $M_R$ evaluations from different groups of experiments
 \cite{lr} one obtains the corresponding lower bounds on $m_{\nu_R}$.

\vbox{
\begin{center}
{\bf Table 3. Lower bounds on the neutrino  masses }

\bigskip

  \noindent
 \begin{tabular}{||l|c|c|c|c|c||} \hline
 Exp.data                  & $M_{W_R}$    & $m_{\nu_R}(\Delta_L)$  &
$m_{\nu_R}$ & $m^3_{\nu_R}(\Delta_L)$ & $m^3_{\nu_R}$ \\ \hline
 $\Delta m_K+$             &   800         & 2.5                    & 2.3
  & 1.44                    &  1.33       \\
 $+B\overline{B}_d+$       &   670          & 2.1                   & 1.9
  & 1.21                    &  1.10        \\
$+b+\beta \beta$           &   740          & 2.31                  & 2.2
  & 1.33                    &  1.26        \\
                           & (450 )        & (1.40 )               & (1.24 )
 & (0.81 )                 &  (0.72 )        \\
man. LR                    & 1400           &  4.4                   & 4.0
    & 2.52                    &  2.3         \\  \hline
 $\Delta m_K+$             & 500            &  1.56                 & 1.43
  &                         &                \\
 $+B\overline{B}_d+$       & 500            &  1.56                 & 1.43
  &                         &                \\
 $+\mu$                    & 740            &  2.30                 & 2.2
  &                         &                 \\
 & (420 )                  &  (1.30 )       & (1.2 )                &
              &                 \\
man. LR                    & 1300           &  4.1                   & 3.8
    &                         &                  \\ \hline
$m_{1R}< 10 MeV$,          & 720            &  2.2                  & 2.0
  &                         &                 \\
Supernova                  & 16200          &  50                    & 46
    &                         &                 \\
Direct search              & 520; 610      &   1.6;1.9
                           & 1.59        &                          &
       \\
Rad.corrections            & 439            &   1.36                 & 1.25
   & 0.80                      & 0.74                 \\ \hline
\end{tabular}
\end{center}      }

 The case of single heavy particle is represented in 2
 central columns of the table, the case of equal masses for
 all 3 generations - in 2 right coloumns. The left
 column of each pair correspond to the theory with $\Delta_L$
 field  while the right one --the theory without
 $\Delta_L$. The experimental
 limits are taken from \cite{lr}.

 One should bear in mind that that all the restrictions on the
 right-handed boson masses  are model dependent.
   However, all these restrictions
 are valid under rather reasonable assumptions and may
 be used for estimations of the right-handed neutrino masses.

       For the theory without left Higgs-Majoron ($\Delta_L$) one
       should bear in mind that that $g_L=g_R$ is only at the GUT scale.
       So $\alpha_{2L}(m_Z)=0.0354$ implies $\alpha_{2R}=0.0265$.
       It is  taken into account in this table.

       The main result of the FT in the Higgs-Majoron sector
is that the absence of quadratic divergences leads to
rather heavy right-handed Majorana neutrinos.

\bigskip

\noindent
{\bf 5.\quad Conclusions.}\\

\medskip

We have shown that in the Standard Model with one and two Higgs
doublets and in the Left-Right symmetric model
the selection rule
based on the vacuum fine-tuning can be implemented for the parameters
of $t$-quark and Higgs-boson
potentials. This is not possible in the original Veltman's formulation.
 To our mind, the approximate
RG invariance is an important property of the fine-tuning conditions
which otherwise do not acquire the universal meaning. Therefore
we have developed the approach different from \cite{18}.
As well we cannot agree with the empirical application of
the original Veltman condition to the physics at the $W$-boson scale
\cite{19} while
it is supposed to be valid at the scale of new physics. It could be
reasonable if this condition was RG-invariant.
In the two-Higgs and Left-Right
symmetric models the vacuum fine-tuning can be realized when
the Yukawa coupling constants for $b$- and $t$-quarks are comparable
that means the hierarchy of v.e.v.'s for the Higgs fields.
In the Left-Right symmetric model the fine-tuning  principles
fix the choice of the representation for the Higgs fields
and could be a useful tool for  the determination
of  possible parameters of the scalar sector of this theory.
Thereby the vacuum fine-tuning may give a resolution between the
different scenarios to generate the hierarchy of quark masses.

\bigskip

We are very grateful to the Organizing Committee of the X Workshop
on QFT and HEP, and especially to Prof. V.I.Savrin,  for many
fruitful discussions and
 for the support of our participation. This report is partially
 supported by the Russian Fond for Basic Research, RFFI Grant
 No. 95-02-05346-a.

 \bigskip

\vspace{5mm}

%\noindent
%{\bf References}\\

\end{document}